\documentclass[journal]{IEEEtran}
\usepackage{cite}
\usepackage{amsmath}
\usepackage{algorithmic}
\usepackage{url}
\usepackage{bbm}
\usepackage{graphicx}
\usepackage{dsfont}
\usepackage{algorithm}
\usepackage{color}
\usepackage{subfigure}
\usepackage{soul}

\begin{document}

\title{Deep Reinforcement Learning Based Mode Selection and Resource Allocation for Cellular V2X Communications}
\author{Xinran~Zhang, Mugen~Peng,~\IEEEmembership{Fellow,~IEEE,} Shi~Yan, and~Yaohua~Sun
\thanks{X. Zhang, M. Peng, S. Yan, and Y. Sun are with the State Key Laboratory of Networking and Switching Technology, Beijing University of Posts and Telecommunications, Beijing 100876, China (e-mail: xrzhang819@bupt.edu.cn; pmg@bupt.edu.cn; yanshi01@bupt.edu.cn; sunyaohua@bupt.edu.cn).}
}
\maketitle

\begin{abstract}
Cellular vehicle-to-everything (V2X) communication is crucial to support future diverse vehicular applications. However, for safety-critical applications, unstable vehicle-to-vehicle (V2V) links and high signalling overhead of centralized resource allocation approaches become bottlenecks. In this paper, we investigate a joint optimization problem of transmission mode selection and resource allocation for cellular V2X communications. In particular, the problem is formulated as a Markov decision process, and a deep reinforcement learning (DRL) based decentralized algorithm is proposed to maximize the sum capacity of vehicle-to-infrastructure users while meeting the latency and reliability requirements of V2V pairs. Moreover, considering training limitation of local DRL models, a two-timescale federated DRL algorithm is developed to help obtain robust model. Wherein, the graph theory based vehicle clustering algorithm is executed on a large timescale and in turn the federated learning algorithm is conducted on a small timescale. Simulation results show that the proposed DRL-based algorithm outperforms other decentralized baselines, and validate the superiority of the two-timescale federated DRL algorithm for newly activated V2V pairs.
\end{abstract}


\section{Introduction}
\IEEEPARstart{T}{o} improve road safety, traffic efficiency, and entertainment experiences on vehicles, vehicle-to-everything (V2X) communication has been recognized as one of
indispensable technologies, which provides wireless connections among vehicles and road infrastructure\cite{bib:vnet1,bib:vnet2}. Up to now, various candidate technical solutions have been proposed, such as cellular V2X and IEEE 802.11p-based dedicated short range communications. Compared to other solutions, cellular V2X is capable of guaranteeing better coverage and quality of service (QoS). In addition,
advanced technologies like non-orthogonal multiple access and millimeter wave communication can be incorporated into cellular V2X to further improve its performance\cite{bib:cv2x, bib:noma}.
Therefore, cellular V2X has drawn much more attention from both industry and academia.

As two vital communication modes in cellular V2X, vehicle-to-infrastructure (V2I) and vehicle-to-vehicle (V2V) communications are exploited to deliver various vehicular applications\cite{bib:app}. On future roads, much more entertainment and traffic-related applications, such as video streaming and crowdsensing, will be undertaken by vehicles, which require frequent access to the Internet or V2X servers via high-capacity V2I communications\cite{bib:v2i_app}. Moreover, safety-critical messages should be forwarded to nearby vehicles in a real-time and reliable manner via V2V communications. For example, as stated in\cite{bib:QoS}, a safety-critical message with the size of 1200 bytes requires the maximum latency of 5 ms and the extreme reliability of 99.999\%. However, it is challenging for existing centralized resource allocation approaches in cellular networks to guarantee such diverse QoS requirements, especially the ultra reliable and low latency requirements.

Motivated by solving the aforementioned challenges, 3GPP has investigated advanced resource allocation approaches for cellular V2X\cite{bib:cv2x}. Firstly, according to the latency and reliability requirements, each vehicular application is endowed with an independent packet priority level. Thus vehicles can prioritize the delivery of safety-critical applications with higher priority levels. Furthermore, novel sensing-based decentralized resource allocation approaches are proposed to guarantee the latency and reliability requirements. Vehicles can sense interference level of each resource block (RB) and then select RBs with lower interference for transmission. Unfortunately, above approaches consider V2V or V2I communications on dedicated resource pool, while severe interference between V2I and V2V communications on shared resource pool is overlooked.

\subsection{Related Works and Chanllenges}
Recently, much attention has been paid to the resource allocation for cellular V2X communication on shared resource pool. Sun \emph{et al.}\cite{bib:c_strom2} propose a cluster-based resource allocation algorithm for V2X communications, where the latency and reliability requirements are transformed into outage constraints that can be tractable with slowly varying large-scale channel information. In \cite{bib:c_ye2}, the ergodic capacity of V2I communications and the reliability of V2V communications are derived based
on statistics of fast fading components, and then centralized resource allocation and power control algorithms are proposed to ensure diverse QoS requirements. Besides, impacts of delayed channel state information (CSI) and queue latency are investigated for cellular V2X communications in \cite{bib:c_delaycsi,bib:c_queue}, respectively. To reduce the pressure on the acquisition of global CSI and computation complexity in above centralized approaches, decentralized approaches are designed for cellular V2X communications. In \cite{bib:d_dqn}, a deep reinforcement learning (DRL) based decentralized resource allocation approach is developed for V2V communications, and each V2V transmitter acts as an autonomous agent who makes decisions based on local observations. Considering mixed centralized/distributed V2X communications, Li \emph{et al.}\cite{bib:cd_select} investigate the joint problem of power control and resource allocation mode selection under different network load conditions. To improve the QoS of vehicles in terms of packet priority and communication link quality, two algorithms are proposed for light and heavy network load scenarios.

For above-mentioned literatures, only V2V communication is employed for the distribution of safety-critical messages among vehicles. However, V2V link actually becomes less reliable when blockage effect is considered, which restricts the performance of V2V communications\cite{bib:cv2x}. To address this issue, a V2I-based forwarding solution can be utilized. In\cite{bib:QoS}, its reliability performance is proven to be enhanced at the cost of higher relay latency and lower spectrum utilization. To guarantee QoS requirements and improve spectrum utilization, communication mode selection and resource allocation should be jointly optimized for cellular V2X communications. Whereas, additional binary mode selection variables make the joint optimization problem intractable with aforementioned optimization algorithms.

As one of the most powerful machine learning tools, reinforcement learning (RL) has recently been applied to the mode selection in wireless networks\cite{bib:mode_q,bib:mode_q2,bib:mode_q3}. Wu \emph{et al.}\cite{bib:mode_q} investigate multi-hop V2I communication and propose a Q-learning based route selection algorithm to realize high throughput and low latency. In\cite{bib:mode_q2}, a distributed approach to mode selection and subchannel allocation for potential device-to-device (D2D) pairs in a D2D enabled cloud radio access network is proposed, in which D2D pairs update their strategies using a RL process. To balance network transmission performance and fronthaul savings in fog computing-based vehicular networks, Yan \emph{et al.}\cite{bib:mode_q3} propose a Q-learning based access mode selection algorithm and a convex optimization based spectrum allocation algorithm.

Nevertheless, multiple sensing components and realistic channel gains generate large-scale continuous state space, which makes Q-learning inefficient. Inspired by \cite{bib:sunbo,bib:ml_dqn}, DRL is capable of addressing above challenges. In DRL, the Q-table is represented by a deep neural network (DNN) and the continuous state can be a direct input to the DNN. Atallah \emph{et al.}\cite{bib:dqn_V} exploit DRL model to learn an optimal transmission mode selection policy from high-dimensional inputs for battery-powered vehicular networks. Considering the highly dynamic topology and time-varying spectrum states in cognitive radio based vehicular networks, a DRL-based optimal data transmission scheduling scheme is designed in\cite{bib:dqn_edi1} to minimize transmission costs while ensuring data QoS requirements. For computation offloading in vehicular netowrks, Zhang \emph{et al.}\cite{bib:dqn_edi2} propose a DRL-based optimal task offloading scheme with varying states of multiple edge servers and multiple vehicular offloading modes.

In above works, the mentioned DRL models are generally trained in a centralized server. In fact, the training data is always distributed at vehicles and unlikely to be uploaded
considering bandwidth overhead and privacy issues. Fortunately, federated learning has the potential to realize distributed learning\cite{bib:fl_edi1,bib:fl_a}.
To achieve high cache efficiency as well as to protect users' privacy, Yu \emph{et al.}\cite{bib:fl_edi1} propose a federated learning based proactive content caching scheme which does not require to gather users' data centrally for training. In \cite{bib:fl_a}, the DRL technique and federated learning framework are integrated to optimize the mobile edge computing, caching and communication. Simulation results show that the proposal has near-optimal performance and relatively low overhead.

Although DRL is promising for the joint mode selection and resource allocation, its application to cellular V2X communication is also faced with several challenges. Firstly, in contrast to assumptions in \cite{bib:d_dqn,bib:mode_actor}, time-varying fast fading channel is always unknown at vehicles due to high dynamics. Besides, to help vehicles make autonomous decisions, decentralized DRL framework is required.
Finally, limited local training data on each vehicle restricts robust learning of DRL model, and improper federated clusters might drastically deteriorate the performance of federated learning.

\subsection{Contributions and Organization}
In this paper, we present a DRL-based decentralized mode\\ selection and resource allocation approach for cellular V2X communications to address the challenges incurred by heterogeneous QoS requirements and unreliable V2V links. The main contributions of this paper are:

\begin{itemize}
\item To alleviate the impacts of unreliable V2V links, a V2I-based forwarding mode is exploited for V2V pairs. Each V2V pair selects either the V2V mode or the V2I  mode based on realistic link qualities. A joint problem of transmission mode selection, RB allocation and power control for cellular V2X communications is formulated to maximize the sum capacity of V2I users while ensuring the latency and reliability requirements of V2V pairs. Different from \cite{bib:mode_actor}, resource sharing among V2V pairs in different transmission modes is considered.

\item We model the formulated problem as a Markov decision process (MDP) and propose a DRL-based decentralized algorithm. Specifically, each V2V pair acts as a DRL agent and makes adaptive decision based on local observations including interference levels, large-scale channel qualities and traffic loads. To guarantee reliability requirement, an effective outage threshold is exploited in the reward function.

\item Considering limited local training data at vehicles, a two-timescale federated DRL-based algorithm is further developed to help obtain robust models. Wherein, a graph-based vehicle clustering is performed to cluster nearby vehicles on a large timescale, while vehicles in the same cluster cooperate to train robust global DRL model through federated learning on a small timescale. Moreover, the global DRL model can be directly downloaded to newly activated V2V pairs, which avoids time-consuming training process.

\item The impacts of vehicular density and outage threshold on the performance are illustrated. Simulation results show that the proposed DRL algorithm outperforms other decentralized algorithms and achieves competitive performance compared to a centralized algorithm. Furthermore, the convergence and superiority of the proposed federated DRL algorithm for newly activated V2V pairs are verified.

\end{itemize}

The remainder of this paper is organized as follows. Section II describes the system model and the formulated optimization problem. Section III presents the basics of DRL and the DRL-based decentralized algorithm. The two-timescale federated DRL-based semi-decentralized algorithm is specified in Section IV and simulation results are illustrated in Section V, followed by the conclusion in Section VI.

\section{System Model and Problem Formulation}
\begin{figure}[!t]
\centering
\includegraphics[width=3.2in]{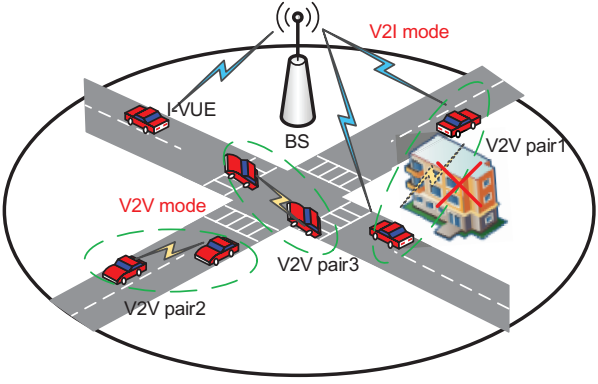}
\caption{Cellular V2X communication in a vehicular network.}
\label{fsystem}
\end{figure}

We consider the cellular V2X communication in a vehicular network which consists of one BS and multiple vehicular user equipments (VUEs), as shown in Fig. \ref{fsystem}. The BS is located at the center of the crossroad, while VUEs are distributed on the roads. Both of them are equipped with a single antenna. Based on active vehicular applications, the whole active VUEs are divided into two parts: $M$ V2I VUEs (I-VUEs) and $K$ V2V pairs. Specifically, I-VUEs upload bandwidth-demanding entertainment information via V2I communication, while each V2V pair contains one V2V receiver and one V2V transmitter which intends to distribute safety-critical messages. According to\cite{bib:QoS}, the maximum frequency of safety-critical messages reaches $10\sim100$ Hz. Thus we assume that the number of V2V pairs is much larger than that of I-VUEs, i.e., $K \gg M$.

Denote the set of I-VUEs as ${\cal M}{\rm{ = }}\left\{ {1,2 \ldots ,M} \right\}$ and the set of V2V pairs as ${\cal K}{\rm{ = }}\left\{ {1,2 \ldots ,K} \right\}$. The total bandwidth is divided into $F$ RBs, denoted by ${\cal F}{\rm{ = }}\left\{ {1,2 \ldots ,F} \right\}$. It is assumed that the number of RBs $F$ is larger than that of I-VUEs $M$ and each I-VUE occupies a single RB for uplink transmission, then there remain $F-M$ unused RBs. To improve the spectrum utilization, each V2V pair can select a single RB from both orthogonal allocated RBs for I-VUEs and remaining unused RBs, and multiple V2V pairs can share the same RB.
For simplicity of notation, we assume RB $m$ is allocated to I-VUE $m$ and use an indicator function $\mathbbm{1}_{m,f} \in \{ 0, 1\}$ to indicate the RB allocation decision for I-VUE.
Specifically, $\mathbbm{1}_{m,f}=1$, if $m=f$; otherwise, $\mathbbm{1}_{m,f}=0$.
For V2V pairs, let $a_{k,f} \in \{ 0, 1\}$ denote whether RB $f$ is allocated to V2V pair $k$, if $a_{k,f} = 1$, then RB $f$ is allocated to V2V pair $k$.

Due to the high mobility of VUEs, we assume that only large-scale channel gain including path loss and shadowing fading is known at the BS and vehicles.
The channel gains from I-VUE $m$ to the BS, from V2V transmitter $k$ to the BS, and between V2V pair $k$ are denoted by ${h_{m,B}}$, ${h_{k,B}}$ and ${h_k}$, respectively. Similarly, we define the interfering channel from V2V transmitter $k$ to the BS, from V2V transmitter $k$ to V2V receiver $j$, and from I-VUE $m$ to V2V receiver $k$ as ${g_{k,B}}$, ${g_{k,j}}$, and ${g_{m,k}}$, respectively.
Considering the blockage of nearby vehicles and buildings, the channel can be in either line of sight (LOS) or non-line of sight (NLOS) state.

\subsection{Communication Modes for I-VUEs and V2V Pairs}
\subsubsection{I-VUEs}
For I-VUEs, only uplink V2I communication is adopted. The uplink signal to interference plus noise ratio (SINR) of I-VUE $m$ is given by
\begin{equation}\label{cvue_sinr}
{\gamma _m^i} = \frac{{{P_m^i}{h_{m,B}}}}{{\sum\limits_{k \in {\cal K}}{\sum\limits_{f \in {\cal F}} {\mathbbm{1}_{m,f}{a_{k,f}}{P_k^v}{g_{k,B}}}  + {\sigma ^2}}}},
\end{equation}
where $P_m^i$ and $P_k^v$ indicate transmit power of I-VUE $m$ and V2V transmitter $k$, respectively.\footnote{In this paper, the superscripts $i$, $v$, $v(I)$, $v(V)$, and $b$ denote
the I-VUE, V2V pair, V2V pair in V2I mode, V2V pair in V2V mode, and BS, respectively.}
$\sigma ^2$ denotes the noise power.
The interference is from the V2V pairs reusing the same RB.

Denote the bandwidth for each RB as $W$, then achievable data rate of I-VUE $m$ can be written as
\begin{equation}\label{cvue_rate}
{R_m^i} = W{\log _2}\left( {1 + {\gamma _m^i}} \right).
\end{equation}

\subsubsection{V2V pairs}
Based on individual channel quality, each V2V pair can select either V2V mode to directly communicate with each other, or V2I mode for indirect communication through the BS. Let ${s_k} \in \{ 0, 1\}$ denote communication mode selection of V2V pair $k$, if $s_k = 1$, V2V pair $k$ chooses the V2I mode; otherwise, V2V pair $k$ selects the V2V
mode. Details of both modes are illustrated below.

In V2V mode, each V2V transmitter directly communicates with its V2V receiver via V2V communication. Interference comes from I-VUEs and V2V pairs which share the same RB. The SINR at V2V receiver $k$ in V2V mode on RB $f$ is given by
\begin{equation}\label{d2d_sinr}
{\gamma ^{v\,(V)}_{k,f}} = \frac{{{a_{k,f}}P_k^v{h_k}}}{{\sum\limits_{m \in {\cal M}} {\mathbbm{1}_{m,f}P_m^i{g_{m,k}} + \sum\limits_{
\scriptstyle j \in {\cal K},\hfill\atop
\scriptstyle j \ne k\hfill } {{a_{j,f}}P_j^v{g_{j,k}} + {\sigma ^2}} } }}.
\end{equation}

Then achievable data rate of V2V pair $k$ in V2V mode can be expressed as
\begin{equation}\label{d2d_rate}
{R^{v\,(V)}_{k}} = \sum\limits_{f \in {\cal F}} {W{{\log }_2}\left( 1 + {\gamma ^{v\,(V)}_{k,f}} \right)}.
\end{equation}

In V2I mode, safety-critical messages are firstly uploaded to the BS and then forwarded to corresponding V2V receivers through downlink. Similar to \cite{forward}, we assume that uplink SINR is smaller than downlink SINR, because the BS has larger transmit power and conducts centralized downlink scheduling.
Therefore, the performance of V2I mode is bounded by uplink SINR. Note that only unused RBs can be allocated to V2V pairs in V2I mode, and each unused RB can be allocated to at most one V2V pair in V2I mode.
The uplink SINR of V2V pair $k$ in V2I mode on RB $f$ is given by
\begin{equation}\label{cellular_sinr}
{\gamma ^{v\,(I)}_{k,f}} = \frac{{{a_{k,f}}P_k^v{h_{k,B}}}}{{\sum\limits_{j \in {\cal K},\,j \ne k} {{a_{j,f}}P_j^v{g_{j,B}}}  + {\sigma ^2}}},
\end{equation}
where the interference is from the V2V pairs which share the same RB and operate in V2V mode.

According to \cite{forward}, achievable data rate of V2V pair $k$ in V2I mode can be expressed as
\begin{equation}\label{cellular_rate}
{R^{v\,(I)}_{k}} \approx \frac{1}{2} \sum\limits_{f \in {\cal F}} {W{\log _2}\left( {1 + {\gamma ^{v\,(I)}_{k,f}}} \right)}.
\end{equation}

\subsection{QoS Requirements of I-VUEs and V2V Pairs}
There are various kinds of vehicular applications with different QoS requirements in vehicular networks. As stated above, I-VUEs undertake bandwidth-demanding entertainment or traffic applications. Thus, QoS requirements of I-VUEs are defined as the minimum capacity requirements to guarantee comfortable experience. In the meantime, V2V pairs should distribute safety-critical messages like cooperative awareness messages in a real-time manner. Any failure of such distributions would threaten road safety.
Therefore, QoS requirements of V2V pairs which deliver these safety-critical messages are the latency and reliability requirements.
The mathematical expression of these QoS requirements are shown as follows.

\subsubsection{Capacity requirements of the I-VUEs}
The capacity requirement of I-VUE $m \in {\cal M}$ is given by
\begin{equation}\label{cellular_qos}
{R^i_m} \ge R_{\text{min} }^i,
\end{equation}
where $R_{\text{min} }^i$ is the minimum capacity requirement of I-VUEs.
For simplicity, we assume that the capacity requirements are the same for all I-VUEs.

\subsubsection{Latency and reliability requirements of the V2V pairs}
The requirements can be divided into two parts: latency requirement and reliability requirement. On the one hand, considering decentralized resource allocation at VUE side, the whole latency for communication between V2V pairs only includes transmission latency, without additional grant-based scheduling latency in the media access control layer. Thus, the latency requirement of V2V pair $k \in {\cal K}$ can be written as
\begin{equation}\label{d2d_latency}
{R_k^v} \ge \frac{{{L_k}}}{{{T_{\text{max} }}}},
\end{equation}
where $L_{k}$ and $T_\text{max}$ are message size in bits and maximum tolerable latency, respectively.
${R_k^v} = (1 - {s_k})R_k^{v\,(V)} + {s_k}R_k^{v\,(I)}$ denotes achievable data rate of V2V pair $k$.

On the other hand, similar to\cite{bib:c_ye2}, we denote outage probability as reliability metric.
With outage threshold $\gamma _{o}$ and tolerable outage probability $p_o$, the reliability requirement of V2V pair $k \in {\cal K}$  is expressed as
\begin{equation}\label{d2d_reliability_old}
\mathds{Pr} \left\{ {{\gamma _{k}^v} \le \gamma _{o}} \right\} \le {p_o},
\end{equation}
where $\gamma _{k}^v = (1 - {s_k})\sum\nolimits_{f\in \cal{F}} {\gamma _{k,f}^{v\,(V)}} + {s_k} \sum\nolimits_{f\in \cal{F}} {\gamma _{k,f}^{v\,(I)}}$ indicates the SINR of V2V pair $k$.
According to \cite{bib:c_ye2}, with Rayleigh fading, reliability constraint \eqref{d2d_reliability_old} can be transformed into
\begin{equation}\label{d2d_reliability}
{\gamma  _k^v} \le \gamma _\text{eff} = \frac{{\gamma _{o}}}{{\ln \left( {\frac{1}{{1 - {p_o}}}} \right)}},
\end{equation}
where $\gamma _\text{eff}$ is the effective outage threshold.
We assume that packet size, maximum tolerable latency, tolerable outage probability are the same for all V2V pairs.

\subsection{Problem Formulation}
In this paper, the global objective is to find the optimal mode selection, RB allocation and power control profile that maximizes the sum capacity of I-VUEs and guarantees the latency and reliability requirements of V2V pairs. The optimization problem is formulated as
\begin{equation}\label{problem}
\begin{aligned}
\mathop {\max }\limits_{{\bf{a,s,p}}} & {\rm{  }}\sum\limits_m {{R_m^i}} {\rm{ }} \\  
\mathrm{s.t.} \; &C1-C3:   \eqref{cellular_qos}\eqref{d2d_latency}\eqref{d2d_reliability}, \\
&C4:{s_k} \in \left\{ {0,1} \right\},\forall k \in {\cal K} \\
&C5:\sum\limits_{f \in {\cal F}} {{a_{k,f}} \le 1} , {a_{k,f}} \in \left\{ {0,1} \right\},\forall k \in {\cal K} \\
&C6:\sum\limits_{k \in {\cal K}} {{s_k}{a_{k,f}} \le 1,} \forall f \in {\cal F} \\
&C7:{P_k^v} \le {P_\text{max}},\forall k \in {\cal K},
\end{aligned}
\end{equation}
where $P_\text{max}$ denotes the maximum transmit power consumption of VUEs.
The optimization objective is to maximize the sum capacity of I-VUEs.
The first three constraints C1-C3 are the capacity requirements of I-VUEs and latency and reliability requirements of V2V pairs.
The fourth constraint C4 indicates each V2V pair can select either V2I mode or V2V mode.
The fifth constraint C5 shows that each V2V pair can be allocated to a single RB and one RB
can be shared by multiple V2V pairs, while the sixth constraint C6 means that each RB can be allocated to at most one V2V pair in V2I mode.
The seventh constraint C7 is to satisfy that transmit power of each V2V pair cannot exceed its maximum value.

The formulated problem \eqref{problem} is a mixed integer nonlinear programming problem which is hard to be directly solved. The reasons are as follows. Firstly, the mode selection indicator $\bf{s}$ and resource allocation indicator $\bf{a}$ are both binary variables, which result in a combinatorial problem. In addition, for the transmission power $\bf{p}$, the optimization object and constraints C1-C3 are non-convex, thus original problem has numerical local optimal solutions\cite{bib:noma}. Recent works for cellular V2X communications mainly focus on centralized approaches\cite{bib:c_strom2,bib:c_ye2}, but the acquisition of global CSI and large computation complexity limit their scalability to dynamic large-scale vehicular networks. Therefore, intelligent decentralized approaches are needed to cope with these challenges.

\section{DRL-based Decentralized Algorithm}
In this section, the basics of RL and recent advances of DRL are elaborated firstly. Then, original problem \eqref{problem} is formulated from the MDP perspective and a DRL-based decentralized algorithm is proposed to solve original problem.

\subsection{Basics of Deep Reinforcement Learning}
As an important branch of machine learning, RL focuses on optimizing action policy and making adaptive decisions by
frequent interaction between time-varying environment and smart agent\cite{bib:RL}.
In general, RL can be modeled as a MDP which is characterized by state space, action space, transition probability and immediate reward.
Based on prior knowledge about transition probability and immediate reward,
RL can be divided into model-based and model-free learning.
Because transition probability and reward are often unknown in realistic environment,
model-free algorithms like Q-learning have drawn much attention.
In Q-learning, action value function is implemented by a Q-table and updated to learn optimal policy.
Hence, Q-learning is suitable for the problem with small-scale and discrete-valued state and action spaces.

To adapt to large-scale dynamic environment, DRL is proposed in \cite{bib:dqn_1} to combine DNNs and RL.
By leveraging non-linear approximation of DNNs\cite{bib:dnn}, Q table is established by a DNN and updates of Q table are transformed into updates of network weights.
Advanced techniques like experience replay and fixed target network have been developed for DRL to accelerate training process and improve convergence performance\cite{bib:sunbo}.
In experience replay technique, DRL models are updated with randomly selected transition histories to break correlations of continuous transition tuples.
And in fixed target network technique, a target Q network is built to predict the target Q value and delayed update of the target Q network is adopted to accelerate and stabilize training process. Therefore, DRL can be applied to large-scale scenario with continuous-valued state space.

Furthermore, actor-critic DRL and multi-agent DRL are proposed to deal with problems with continuous-valued action spaces and multiple learning agents, respectively. More details related to these approaches can be found in\cite{bib:ml_dqn,bib:marl}.
Note that the considered problem in this paper is with continuous state spaces and discrete action space, therefore a deep Q-network (DQN) based DRL framework is exploited.

\subsection{DRL-based Decentralized Algorithm}
The latency and reliability requirements of V2V pairs, resource sharing among V2V pairs and unreliable V2V links introduce much complexity in the aspects of interference control, continuous-value state space and large action space.
Inspired by core idea of DRL, we transform original problem \eqref{problem} into a MDP. As shown in Fig.\ref{figure:DRL}, the framework of DRL consists of DRL agents and cellular V2X environment interacting with each other. Each V2V pair is considered as an intelligent DRL agent performing local decision.
In cellular V2X, time is divided into subframes denoted by $\{ 0,1,...,t,...\}$. The scheduling period of V2V pairs can be an arbitrary positive integer.
Without loss of generality, the scheduling period is set as 1 subframe for the whole V2V pairs.
Three key elements of the MDP model, i.e., state space, action space and immediate reward, are defined as follows:

\begin{figure}[!t]
\centering
\includegraphics[width=3.3in]{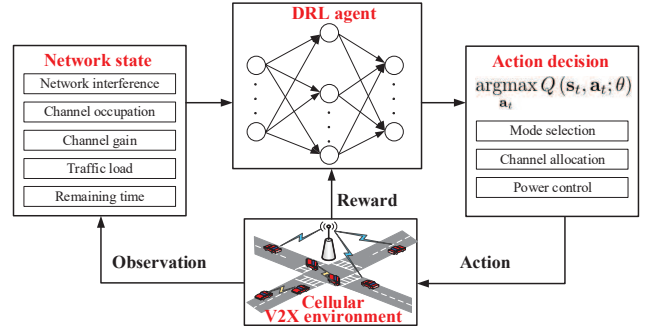}
\caption{A detailed framework of deep reinforcement learning.}
\label{figure:DRL}
\end{figure}

\noindent $\bullet$  \textbf{ State Space}

For each V2V pair, the observed state $\mathbf{s}_t$ at subframe $t$ consists of
seven parts: the received interference power at the V2V receiver and the BS on each RB at previous subframe ${{\bf{I}}^v_{t - 1}} = \left\{ {{I^v_{1,t - 1}},{I^v_{2,t - 1}}, \ldots ,{I^v_{F,t - 1}}} \right\}$, ${{\bf{I}}^b_{t - 1}} = \left\{ {{I^b_{1,t - 1}},{I^b_{2,t - 1}}, \ldots ,{I^b_{F,t - 1}}} \right\}$,
 the number of selected neighbors on each RB at previous subframe ${{\bf{N}}_{t - 1}}= \left\{ {{N_{1,t - 1}},{N_{2,t - 1}}, \ldots ,{N_{F,t - 1}}} \right\}$ , the large-scale channel gains from the V2V transmitter to its corresponding V2V
receiver and the BS ${{{h }_{k,t}}}$, ${{{ h }_{k,B,t}}}$, current load ${L_t^r}$ and remaining time to meet the latency threshold ${T_t^r}$.
Thus, the state can be described as
\begin{equation}\label{state1}
{\mathbf{s}_t} = \left\{ {{{\bf{I}}^v_{t - 1}},{{\bf{I}}^b_{t - 1}},{{\bf{N}}_{t - 1}},{{ h }_{k,t}},{{h }_{k,B,t}},L_t^r,T_t^r} \right\}.
\end{equation}
The state space can be expressed as $\mathcal{S} = \{\mathbf{s}^i| i = 1,2,...\}$, where $\mathbf{s}^i$ is potential state $i$.

\noindent $\bullet$  \textbf{ Action Space}

The action of each V2V pair is defined as $\mathbf{a}_t = \left\{ {{a},{s},{p}} \right\}$. Consistent with notation in Section II,
$a\in {\cal F}$, $s\in \{0, 1\}$, and $p\in \{0,\frac{1}{N_p-1}P_\text{max},\frac{2}{N_p-1}P_\text{max},...,P_\text{max}\}$ represent the RB allocation, communication mode selection, and transmit power level of the V2V transmitter, respectively.
Note that, we adopt discrete power control scheme \cite{distrete_pc} and assume transmit power of VUEs  has $N_p$ levels. Thus, the size of action space $\mathcal{A}$ is $2FN_p$.

\noindent $\bullet$  \textbf{ Immediate Reward}

In order to maximize the sum capacity of I-VUEs and guarantee the QoS requirements of both I-VUEs and V2V pairs,
the immediate reward at subframe $t$ is defined as
\begin{equation}\label{reward}
\begin{split}
r_t = & \sum\limits_{m \in {\cal M}} {{c_1}{R_m^i}}  + \sum\limits_{m \in {\cal M}} {{c_2}G\left( {{R_m^i} - R_{\text{min} }^i} \right)}  + \\
&\sum\limits_{k \in {\cal K}} {{c_3}G\left( {{\gamma _k^v} - \gamma _\text{eff}^v} \right)}  + \sum\limits_{k \in {\cal K}} {c_4}G\left({R_k^v - \frac{L^r_t}{T^r_t}} \right).
\end{split}
\end{equation}
Here, $G(x)$ is a piecewise function
\begin{equation}\label{function}
G\left( x \right) = \left\{ \begin{array}{l}
A,x \ge 0\\
x,x < 0,
\end{array} \right.
\end{equation}
where $A > 0$ is set as a positive constant to indicate revenue.

The immediate reward \eqref{reward} is composed of four parts. The first parts corresponds to the sum capacity revenue of I-VUEs, while the second part
indicates penalty of unsatisfied capacity for I-VUEs.
The third and fourth parts denote impacts of the reliability and latency requirements.
And $c_1$, $c_2$, $c_3$, $c_4$ are weights of each part to balance the revenue and penalty.

At the beginning of subframe $t$, each V2V pair observes their own state $\textbf{s}_t$ and then performs joint mode selection and
resource allocation $\mathbf{a}_t$ based on established action value function $Q\left( \textbf{s}_t, \textbf{a}_t;\theta\right)$.
The action value function is defined as
\begin{equation}\label{Qfuntion}
{Q }\left( {\mathbf{s},\mathbf{a};\mathbf{\theta}} \right) = \mathds{E}\left[ {\sum\limits_{t' = t}^T {{\gamma ^{t' - t}}{r_{t'}}}|{\mathbf{s}_t} = \mathbf{s},{\mathbf{a}_t} = \mathbf{a};\theta } \right],
\end{equation}
where $T$ and $0 < \gamma < 1$ are the terminal step of each epoch and discount factor that represents the impact of future reward, respectively.

Afterwards, based on actions taken by different agents, the cellular V2X environment transits to a new state $\mathbf{s}_{t+1}$
and the agents gather the immediate reward $r_t$ from the environment.
Specifically, the V2V pairs broadcast their remaining load and experienced SINR
to nearby V2V pairs; and I-VUEs broadcast their experienced data rate.
Based on above elements, each V2V pair computes the immediate reward with equation \eqref{reward}.

With $r_t$ and $\mathbf{s}_{t+1}$, V2V pairs can update the weights of DQNs by minimizing loss function $L\left( \theta  \right)$ at each step.
Similar to \cite{bib:dqn_1}, mean square error is adopted as the loss function, i.e.,
\begin{equation}\label{loss}
L\left( \theta  \right) = \mathds{E}\left\{ {\left( y_t - { Q\left( {{\mathbf{s}_t},{\mathbf{a}_t};\theta } \right) }   \right)^2} \right\},
\end{equation}
where $ y_t = r_t + \gamma \mathop {{\rm{max}}}\limits_{{\mathbf{a}_{t + 1}}} \hat{Q}  \left( {{\mathbf{s}_{t + 1}},{\mathbf{a}_{t + 1}};{\theta ^ - }} \right)$.
Here,
$\hat{Q} \left( {{\mathbf{s}},{\mathbf{a}};{\theta ^ - }} \right)$ is the target Q network updated every $N_Q$ steps.

The DRL procedures of solving original problem can be concluded in \textbf{Algorithm \ref{alg1}} in which the experience replay and fixed target network techniques are considered.
Note that, the $\epsilon$ greedy policy means that the agent randomly selects an action $\mathbf{a}_t \in \mathcal{A}$ with a probability of $\epsilon$,
and chooses the optimal action $\mathbf{a}_t = \mathop {{\rm{argmax}}}\limits_{{\mathbf{a}}} Q  \left( {{\mathbf{s}}_{t},{\mathbf{a}};{\theta }} \right) $
with a probability of $1-\epsilon$. Here, $\epsilon$ is the exploring factor.

\begin{algorithm}[htb]
\caption{DRL-based decentralized algorithm}
\label{alg1}
\begin{algorithmic}[1]
\STATE \textbf{Input:} Discount factor $\gamma$, learning rate $\beta$, replay capacity $N_\text{memory}$, and batch size $B$.
\STATE \textbf{Initialization} Initialize a DNN with random weights $\bf{\theta}$ as the action value function $Q\left( {{\mathbf{s}},{\mathbf{a}};\theta}\right)$,
and make a copy of it to represent the target action value function ${\hat{Q}  }\left( {{\mathbf{s}},{\mathbf{a}};\theta^ -}\right)$. Then, V2V pairs randomly select actions
until storing $N$ transitions in the replay memory.
\STATE \textbf{For} epoch $e=1,...E$: \\
\STATE\qquad Observe the state $\mathbf{s}_1$.\\
\STATE\qquad \textbf{For} step $t=1,...T$:\\
\STATE\qquad \qquad V2V pairs select action $\mathbf{a}_t$ according to $\epsilon$ greedy\\\qquad \qquad policy.\\
\STATE\qquad \qquad Obtain current reward $r_t$ and next state $\mathbf{s}_{t+1}$,\\\qquad \qquad then store transition tuples $(\mathbf{s}_t,\mathbf{a}_t,r_t,\mathbf{s}_{t+1})$ in \\\qquad \qquad  the replay memory.\\
\STATE\qquad \qquad Randomly sample a mini-batch of transition\\\qquad \qquad tuples from the reply memory, and perform a\\\qquad \qquad gradient descent step on \eqref{loss} with respect to \\\qquad \qquad network weights $\theta$.\\
\STATE\qquad \qquad Every $N_Q$ steps update $\theta^ - = \theta$.\\
\STATE\qquad \textbf{End For}\\
\STATE\textbf{End For}\\
\end{algorithmic}
\end{algorithm}

\section{Federated DRL-based Semi-Decentralized Algorithm}
Although approximate optimal solution can be derived by the proposed DRL-based decentralized algorithm,
the stringent latency requirement and lack of training data pose huge challenges to the training of accurate DRL models.
Furthermore, without well-trained DRL models, newly activated V2V pairs might make inferior local decisions
and degrade global performance.
Finally, well-trained DRL models can be easily outdated due to high mobility of vehicles.
Considering nearby V2V pairs often experience similar channel quality and environment observations,
they can be employed to train robust DRL models.

In this section, a two-timescale federated DRL framework is proposed to train robust DRL models and
improve the performance of newly activated V2V pairs.
Specifically, the centralized VUE clustering on a large timescale and federated DRL on a small timescale are elaborated.

\subsection{Two-Timescale Federated DRL Framework}
In this subsection, the basics of federated learning and its application are firstly illustrated.
Afterwards, a two-timescale federated DRL framework is designed to overcome the aforementioned challenges.

\subsubsection{Basics of federated learning}
Although great breakthrough has been made in the areas of DNN and DRL,
existing data computation and model training are more likely to be performed in a centralized server or a computer cluster.
However, due to privacy and communication cost issues, most devices are not willing to share private data.
On the other hand, the model training and data analysis of local data at the device side are always time-consuming and imprecise.
To cope with these challenges, federated learning is proposed to allow a loose federation of participating devices
with the coordination of a central server\cite{bib:fl}.

The core idea of federated learning is to decouple model training from the need for direct access to raw training data.
By leveraging local training based on local raw data at device side and infrequent averaging of local models at centralized
server, federated learning can effectively enhance the training performance of distributed DNN and DRL.
As for communication cost, the uploading overheads of federated learning is negligible compared to that of centralized learning, as evaluated in\cite{bib:fl_a}. The reasons are two-folds.
On the one hand, by utilizing advanced model compression techniques\cite{bib:fl_eff}, the size of uploading models in federated learning is smaller than that of raw data sets.
On the other hand, the averaging period is much larger than the training period.
Recently, federated learning has been introduced to mobile edge networks and integrated with DRL to perform intelligent distributed resource allocation\cite{bib:fl2}.

\subsubsection{Two-timescale federated DRL framework}
\begin{figure}[!t]
\centering
\includegraphics[width=3.2in]{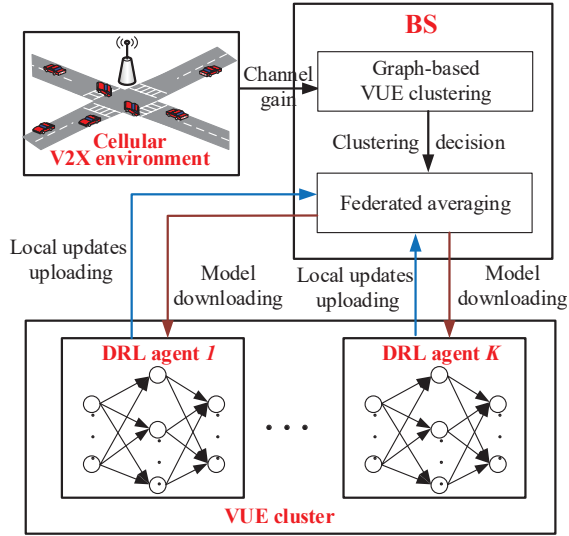}
\caption{A detailed framework of federated deep reinforcement learning.}
\label{figure:FL}
\end{figure}

Inspired by above works, we propose an integral framework of federated learning and DRL, denoted as federated DRL,
for the mode selection and resource allocation in vehicular networks.
Fig. \ref{figure:FL} shows the schematic diagram of the proposed two-timescale federated DRL framework. The whole process can be divided into two procedures in different timescales.

On the large timescale, the BS periodically constructs undirected graphs based on the large-scale channel gains, and groups nearby VUEs with the similar channel gains.
In addition, the candidate RB group is determined for each cluster to reduce network dimension and the probability of resource collision.

On the small timescale, federated learning is introduced to average local models of V2V pairs in the same cluster.
Specifically, V2V pairs in the same cluster asynchronously select their actions and train local models in each subframe.
Every a few hundreds of subframes, the local models of member V2V pairs in the same cluster are uploaded and averaged,
and then the resulting global network is feedback to the whole member V2V pairs.
In particular, the global network can be downloaded to newly activated V2V pairs to avoid time-consuming training process.

\subsection{Centralized VUE Clustering on a Large Timescale}
The details of graph-based VUE clustering are illustrated below.
Firstly, we construct an undirected graph $G\left( {V,E} \right)$ in which each V2V pair or I-VUE is modeled as a vertex and two vertices are joined by an edge.
Here, $V\left( {G} \right)$ and $E\left( {G} \right)$ denote the sets of vertices and edges, respectively.
Note that in vehicular networks, the link between nearby VUEs is unreliable due to blockage, thus we adopt large-scale channel gains rather
than Euclidean distances as the weights of edges.
Considering the worst cases, the weight of the edge between vertex $i$ and $j$ is defined as follows
\begin{equation}\label{weight}
{w_{j,k}} = \max \left\{ {{{g }_{j,k}},{{g }_{k,j}}} \right\},\forall j \ne k.
\end{equation}

In order to cluster nearby VUEs with similar channel gains,
the clustering problem is transformed into a graph partition problem with the aim to maximize the sum weights of edges inside clusters, i.e.,
\begin{equation}\label{cluster_problem}
\begin{aligned}
\mathop {\max }\limits_{{\cal {C}}_1,...,{\cal {C}}_C}\quad & \sum\limits_{c = 1}^C {\left( {\sum\limits_{i,j \in {{\cal {C}}_c}} {{w_{i,j}}} } \right)} \\
\mathrm{s.t.} \qquad &C1:{{\cal {C}}_1} \cup {{\cal {C}}_2} \cup  \cdots  \cup {{\cal {C}}_C} = V\left( G \right), \\
&C2:{{\cal {C}}_i} \cap {{\cal {C}}_j} = \emptyset ,\forall i \ne j,
\end{aligned}
\end{equation}
where $C$ and ${\cal {C}}_c$ denote the predefined number of clusters and the $c$th cluster set, respectively.
As shown in \cite{bib:c_strom2}, this graph partitioning problem is NP-hard.
Traditional Euclidean distance-based clustering approaches like $K$-means and $K$-medoids are not applicable,
because weights in constructed undirected graph is not based on Euclidean distance.
Moreover, heuristic approaches developed in \cite{bib:c_strom2,bib:c_ye2} are dependent on random initialization of each cluster.
To cope with above issues, we adopt spectral clustering to solve the problem.
In spectral clustering, similarity-based weights are exploited and the optimal solution is obtained through multiple searches.
The detailed description of spectral clustering can be found in \cite{bib:sc}.

In order to mitigate interference among VUEs, V2V pairs and I-VUEs in the same cluster should be allocated with orthogonal resources.
Therefore, based on the clustering results, the candidate RB group for cluster $C_c$ is defined as $\mathcal{F}_c =  \mathcal{F}/ {\left\{ {m|m \in \mathcal{M},m \in {\mathcal{C}_c}} \right\}}$.

The centralized VUE clustering algorithm on a large timescale is concluded in\textbf{ Algorithm \ref{alg3}}.

\begin{algorithm}[htb]
\caption{ Centralized VUE clustering algorithm on a large timescale}
\label{alg3}
\begin{algorithmic}[1]
\STATE \textbf{For} clustering period $ n = 1, 2,...:$
\STATE\qquad Initialize the undirected graph $G\left( {V,E} \right)$.
\STATE\qquad Calculate the cluster sets $C_1,...,C_C$ by using spectral\\\qquad clustering method\cite{bib:sc}.
\STATE\qquad \textbf{For} each cluster $ c = 1, 2,....:$
\STATE\qquad\qquad Determine the candidate RB group with $\mathcal{F}_c = $ \\\qquad\qquad $ \mathcal{F}/ {\left\{ {m|m \in \mathcal{M},m \in {\mathcal{C}_c}} \right\}}$.
\STATE\qquad \textbf{End For}
\STATE\textbf{End For}
\end{algorithmic}
\end{algorithm}

\subsection{Federated DRL on a Small Timescale}
With the cluster sets and candidate RB groups obtained by\textbf{ Algorithm \ref{alg3}},
federated learning could be introduced to help train robust DRL models.
The whole process of federated DRL can be divided into numerous coordination rounds.
At the beginning of each coordination round $r=1,2,...$, the BS distributes pre-trained or averaged model to the V2V pairs in the same clusters.
Then each V2V pair performs \textbf{Algorithm \ref{alg1}} to train their own models based on local training data.
Until next round, the BS selects the V2V pairs from the same cluster to upload their models, performs federated averaging,
and then re-distributes the averaged model back.

The core process of federated DRL is federated averaging.
Here, we adopt mini-batch based stochastic gradient descent for federated averaging.
With $N_c^v$ V2V pairs in cluster set $c$, the weights of the global model can be updated by
\begin{equation}\label{fed_avg}
{\theta _{r + 1}} \leftarrow \sum\limits_{k \in {\mathcal{C}_c}} {\frac{{B^k}}{B}\theta _{r+1}^{k}},
\end{equation}
where $\theta _{r + 1}$ and $\theta _{r + 1}^k$ are the weights of global Q network and local Q network at V2V pair $k$ on round $r + 1$, respectively.
$B$ and $B^k$ are the sum batch size for all V2V pairs and the training batch size of V2V pair $k$.
Note that \eqref{fed_avg} is equal to ${\theta _{r + 1}} \leftarrow {\theta _r} - \beta \sum\nolimits_{k \in {C_c}} {\frac{{B^k}}{B}{\nabla L_k}\left( {{\theta _r}} \right)}$,
where ${\frac{{B^k}}{B}{\nabla L_k}\left( {{\theta _r}} \right)}$ is the gradient with respect to ${\theta _r}$\cite{bib:fl}.

In federated DRL, each V2V pair independently selects its own action based on local observations, without any knowledge of actions selected by other V2V pairs.
As a result, the observations of each V2V pairs cannot characterize the whole environment and resource collision in the same cluster will severely degrade the performance.
In order to mitigate above issues, an asynchronous scheme is introduced in the federated DRL-based algorithm.
Specifically, the whole discrete subframes are divided into multiple subframe blocks, and subframe block $c$ consists of $N_c^v$ subframes.
Each V2V pair in the same cluster set is allocated to a specific subframe and asynchronously performs action selection at the allocated subframe.

For newly activated V2V pairs, they request the BS to decide cluster set which they belong to.
Then, the global DRL model and detailed network parameters of their specific clusters are downloaded to these newly activated V2V pairs.
In this way, time-consuming training process of local DRL models could be avoided.

Finally, the federated DRL-based algorithm is concluded in\textbf{ Algorithm \ref{alg2}}.

\begin{algorithm}[htb]
\caption{ Federated DRL-based semi-decentralized algorithm on a small timescale}
\label{alg2}
\begin{algorithmic}[1]
\STATE The BS initializes the Q network with $\theta$, and distributes the Q network to V2V pairs in the scenario.\\
\STATE\textbf{For} each coordination round $r=1,2,...:$\\
\STATE\qquad\textbf{For} each cluster $ c = 1, 2,....:$
\STATE\qquad \qquad V2V pairs in the cluster perform \textbf{Algorithm 1} \\\qquad \qquad in an asynchronous manner at each subframe.\\
\STATE\qquad\qquad  Upload local model weights $\theta _{r}^k$ to the BS.\\
\STATE\qquad\qquad  The BS calculates the global model weights by \\\qquad \qquad using federated averaging with \eqref{fed_avg}, and distri- \\\qquad \qquad butes this global model to all V2V pairs in the \\\qquad \qquad cluster.\\
\STATE\qquad\textbf{End For}
\STATE\qquad\textbf{If} there are newly activated V2V pairs:
\STATE\qquad\qquad  Request the BS to calculate its cluster by using \\\qquad \qquad  \textbf{Algorithm 2}.
\STATE\qquad\qquad  The BS distributes the global model of the speci- \\\qquad \qquad fic cluster to the newly activated V2V pair.\\
\STATE\qquad\qquad  The V2V pair performs \textbf{Algorithm 1} in an asyn- \\\qquad \qquad chronous manner.
\STATE\qquad\textbf{End If}
\STATE\textbf{End For}
\end{algorithmic}
\end{algorithm}

\section{Simulation and Numerical Results}
In this section, the performances of the proposed DRL and federated DRL algorithms for cellular V2X communications are evaluated through simulations.

Similar to the assumptions in \cite{bib:mode_actor}, we consider a crossroad scenario in the simulation
where vehicles are dropped in the crossroad based on spatial Poisson process and a BS is located at the center.
The crossroad size is $1$ km $\times 1$ km and each road consists of two lanes in each direction.
$5$ active I-VUEs and $K$ active V2V transmitters are randomly selected among the vehicles, and each V2V transmitter
builds a V2V link with the farthest vehicle in its broadcast range.
The determination of LOS status, path loss, shadowing and fast fading parameters is based on the
urban street scenario in 3GPP TR 37.885 \cite{bib:QoS}.
As defined in \cite{bib:QoS}, the latency and reliability requirements for safety-critical messages of $800$ bytes
are $10$ ms and $99\%$ with the outage threshold $3$ dB, respectively.
The capacity requirement of the I-VUEs is $3$ bps/Hz.
The predefined number of the clusters is set as $5$.
Throughout the simulations, unless otherwise specified, we adopt the parameters reported in Table \ref{table-asp1}.

\begin{table}[!t]
\renewcommand{\arraystretch}{1.3}
\caption{Default Simulation Parameters}
\label{table-asp1}
\centering
\begin{tabular}{|l|l|}
\hline
\textbf{Parameter  }                    & \textbf{Value}                                                   \\ \hline
Carrier frequency              & 2 GHz                                                   \\ \hline
Number of RBs                  & 10                                                      \\ \hline
Bandwidth of each RB           & 180 kHz                                                 \\ \hline
Number of I-VUEs               & 5                                                       \\ \hline
Number of V2V pairs            & 5, 10, 15, 20                                     \\ \hline
Path loss model of V2V links   & \begin{tabular}[l]{@{}l@{}}LOS: $44.23 + 16.7 {\log _{10}}\left( d \right)$\\ NLOS: $42.52 + 30.0 {\log _{10}}\left( d \right)$\end{tabular} \\ \hline
Path loss model of V2I links   & \begin{tabular}[l]{@{}l@{}}LOS: $38.40 + 21.0 {\log _{10}}\left( d \right)$\\ NLOS: $38.40 + 31.9 {\log _{10}}\left( d \right)$\end{tabular}  \\ \hline
Vehicle velocity               & 36 km/h                                                 \\ \hline
Distance between V2V pairs     & 150 m                                                   \\ \hline
Maximum transmit power of VUEs & 23 dBm                                                  \\ \hline
Antenna configuration          & 1 antenna for VUE and BS                                \\ \hline
Noise power                    & -114 dBm                                                \\ \hline
\end{tabular}
\end{table}

\begin{table}[!t]
\renewcommand{\arraystretch}{1.3}
\caption{Simulation Parameters for DRL}
\label{table-asp2}
\centering
\begin{tabular}{|c|c|}
\hline
\textbf{Parameter }                      & \textbf{Value}          \\ \hline
Learning rate                   & 0.001          \\ \hline
Discount factor                 & 0.70           \\ \hline
Initial exploration             & 1              \\ \hline
Final exploration               & 0.01           \\ \hline
Total exploration steps         & 1000           \\ \hline
Replay memory size              & 3000           \\ \hline
Minibatch size                  & 8              \\ \hline
Network update frequency        & 2              \\ \hline
Target network update frequency & 30             \\ \hline
Federated averaging frequency   & 100             \\ \hline
Number of steps in each epoch   & 10             \\ \hline
Weights in reward function       & 0.1, 0.9, 1, 1 \\ \hline
\end{tabular}
\end{table}

The adopted DQN in the simulation is a fully connected neural network constructed by an input layer, an hidden layer and an output layer.
The number of neurons in the hidden layer is 256, while the ReLu and adaptive moment estimation method are utilized as the activation function
and optimizer, respectively.
All other parameters related to the DQN are listed in Table \ref{table-asp2}.
Note that the listed parameters are selected from multiple simulation tests to balance complexity and performance of DRL algorithm.

To verify the efficiency of our proposals, three algorithms are adopted in our simulation study:
\begin{itemize}
  \item \textbf{Centralized algorithm} \cite{bib:c_ye2}: In this algorithm, the transmission mode is determined by a greedy scheme,
  then the optimal RB and transmit power are allocated to each V2V pair based on the Hungarian algorithm and closed-form solution in \cite{bib:c_ye2}. Note that, the BS is assumed to possess global CSI and perform this algorithm in a centralized manner.
  \item \textbf{DRL-based algorithm without mode selection} \cite{bib:d_dqn}: In this algorithm, only V2V mode is adopted for V2V pairs.
  Each V2V pair independently selects its RB and transmit power based on local DRL model.
  \item \textbf{Random selection algorithm}\cite{bib:cv2x}: In this algorithm, the V2V pair randomly selects the transmitting RB from a candidate RB pool which consists of $5$ RBs with lower interference. The transmit power is set as the maximum transmit power and only V2V mode is adopted.
\end{itemize}

\subsection{Network Performance versus the Number of V2V pairs}
\begin{figure}[!t]
\centering
\includegraphics[scale=0.6]{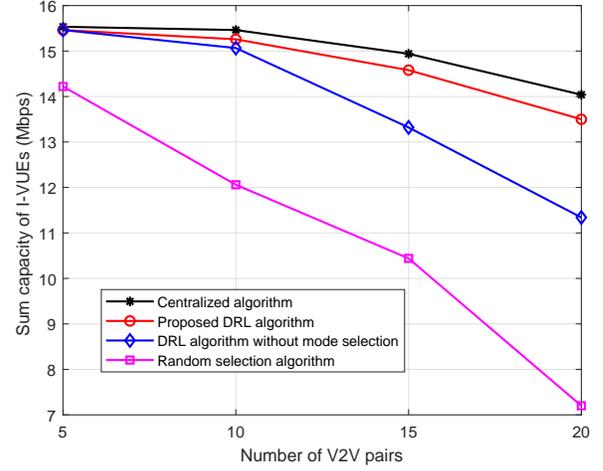}
\setlength{\belowcaptionskip}{-100pt} \vspace*{-10pt}
\caption{Sum capacity of I-VUEs versus the number of V2V pairs.}
\label{RATE1}
\end{figure}

\begin{figure}[!t]
\centering
\includegraphics[scale=0.6]{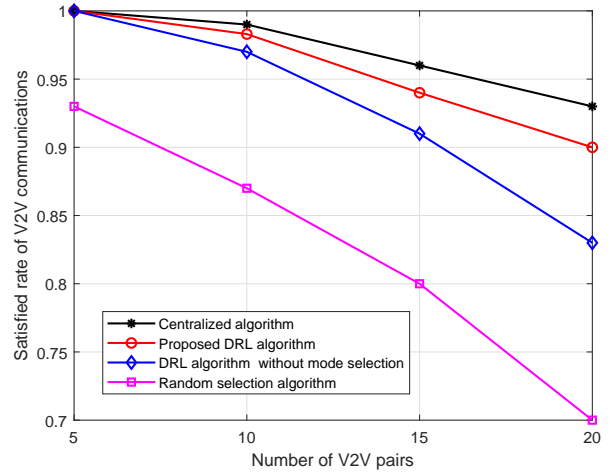}
\setlength{\belowcaptionskip}{-100pt} \vspace*{-10pt}
\caption{Satisfied rate of V2V communications versus the number of V2V pairs.}
\label{PRR1}
\end{figure}

Fig. \ref{RATE1} and Fig. \ref{PRR1} show the sum capacity of I-VUEs and the satisfied rate of V2V communications versus different numbers of V2V pairs. We can see that the proposed DRL algorithm outperforms other decentralized algorithms from the perspectives of both performance metrics.
This is because as the number of V2V pairs increases, more V2V links are in NLOS state due to the blockage of nearby vehicles.
The proposed DRL algorithm can identify these unstable V2V links and select best transmission mode based on local observations, while other decentralized algorithms fail.
In addition, when V2V pairs in the crossroad select the V2I mode, lower transmit power is needed to guarantee the reliability performance,
which reduces the whole interference level, especially when the number of V2V pairs is large.
As a result, the difference between the proposed DRL algorithm and other decentralized algorithms increases with more V2V pairs.
Moreover, it can be further observed that the proposed DRL algorithm achieves close performance to that of centralized algorithm.
Note that the acquisition of global CSI, high computation complexity and frequent rescheduling for arbitrary activated V2V pairs
make the centralized algorithm inefficient in large-scale vehicular networks.
However, in our proposed DRL algorithm, each V2V pair makes decentralized decision based on only local observations, meanwhile the well-trained DRL model can be transferred to
newly activated V2V pairs by transfer learning or federated learning.

\subsection{Network Performance versus Outage Threshold}
\begin{figure}[!t]
\centering
\includegraphics[scale=0.6]{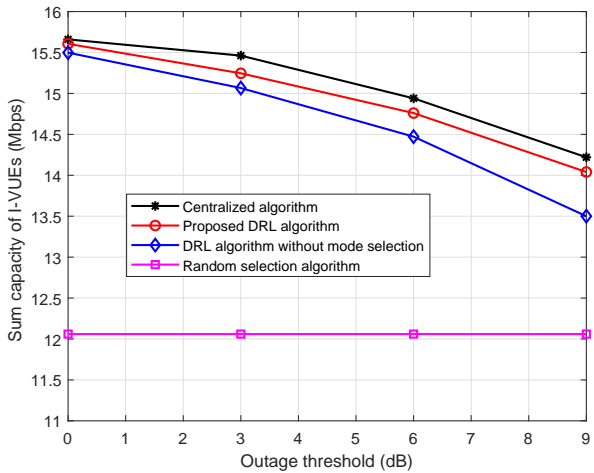}
\setlength{\belowcaptionskip}{-100pt} \vspace*{-10pt}
\caption{Sum capacity of I-VUEs versus outage threshold.}
\label{RATE2}
\end{figure}

\begin{figure}[!t]
\centering
\includegraphics[scale=0.6]{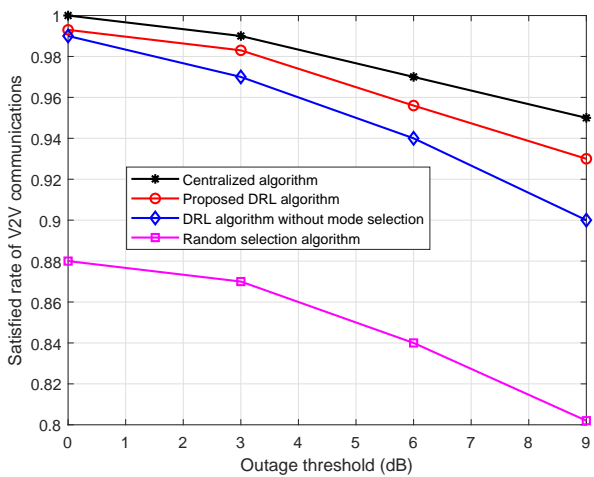}
\setlength{\belowcaptionskip}{-100pt} \vspace*{-10pt}
\caption{Satisfied rate of V2V communications versus outage threshold.}
\label{PRR2}
\end{figure}

Fig. \ref{RATE2} and Fig. \ref{PRR2} show the sum capacity of I-VUEs and the satisfied rate of V2V communications versus different outage thresholds. The number of V2V pairs is set as 10.
It can be observed that with the increase of outage threshold, the sum capacity of I-VUEs and satisfied rate of V2V communications decline for the centralized algorithm, proposed DRL algorithm and DRL algorithm without mode selection.
In addition, the proposed DRL algorithm outperforms other decentralized algorithms.
This is because that with larger outage threshold, V2V pairs tends to select larger transmission power level to guarantee the reliability requirement \eqref{d2d_reliability},
which results in more severe interference to nearby I-VUEs and V2V pairs.
The proposed DRL algorithm can effectively alleviate the interference by adaptively selecting optimal transmission mode.
Note that in random selection algorithm, the V2V pair randomly selects its transmitting RB without considering the capacity of I-VUEs, thus the sum capacities of I-VUEs are the same for different outage thresholds.

\subsection{Effectiveness of the Federated DRL Algorithm}

\begin{figure}[!t]
\centering
\includegraphics[scale=0.6]{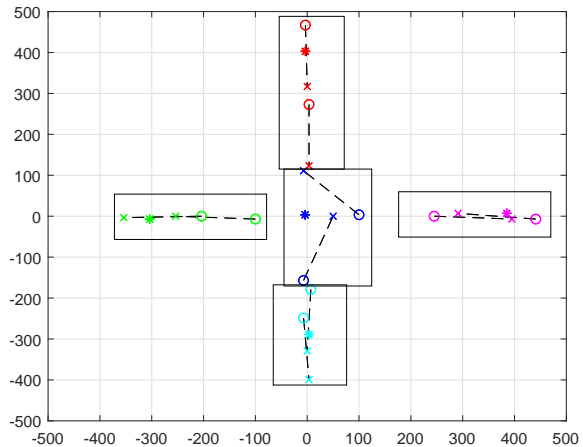}
\setlength{\belowcaptionskip}{-100pt} \vspace*{-10pt}
\caption{Cluster result of proposed graph-based clustering algorithm when the number of V2V pairs is 10.  (o, x, and * represent V2V transmitter, V2V receiver, and I-VUE, respectively.)}
\label{cluster_result}
\end{figure}

Fig. \ref{cluster_result} shows the cluster result of the proposed centralized VUE clustering algorithm.
The number of V2V pairs is set as $10$. Note that in our simulation, the channels between VUEs in different streets are in NLOS state.
It can be observed that our proposed VUE clustering algorithm can identify neighboring V2V pairs and I-VUEs with similar LOS states and put them into the same cluster.
This is because large-scale channel gain is utilized as the weights of edges in our constructed graph and spectral clustering is adopted.

\begin{figure}[!t]
\centering
\includegraphics[scale=0.6]{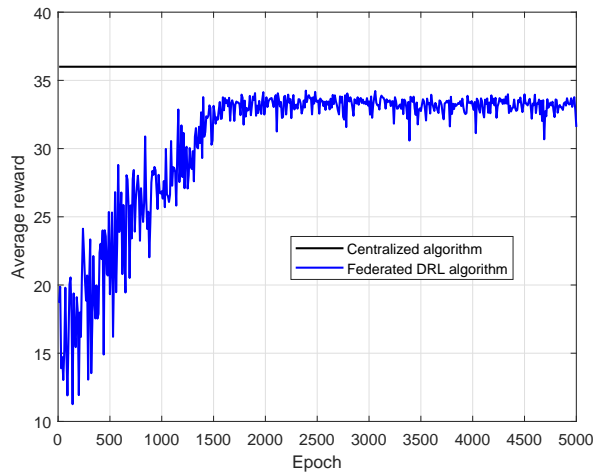}
\setlength{\belowcaptionskip}{-100pt} \vspace*{-10pt}
\caption{Learning process of the federated DRL algorithm.}
\label{CONVERGE1}
\end{figure}

Fig. \ref{CONVERGE1} shows the learning process of the federated DRL algorithm. The number of V2V pairs is set as $10$.
It can be observed that the average reward in the proposed federated DRL algorithm is low at the beginning of learning process. With the increase of epoch, the average reward increases until it reaches a relatively stable value.
This shows the convergence performance of the proposed federated DRL algorithm.
We can also observe that the convergent average reward in the federated DRL algorithm is close to the optimal reward in centralized algorithm.
Furthermore, it should be noted that the proposed DRL algorithm achieves similar convergence performance with federated DRL algorithm in our simulation.

To verify the effectiveness of well-trained federated DRL model for newly activated V2V pairs,
we consider a scenario where the original number of V2V pairs is $10$ and a V2V pair is newly activated.

\begin{figure}[!t]
\centering
\includegraphics[scale=0.6]{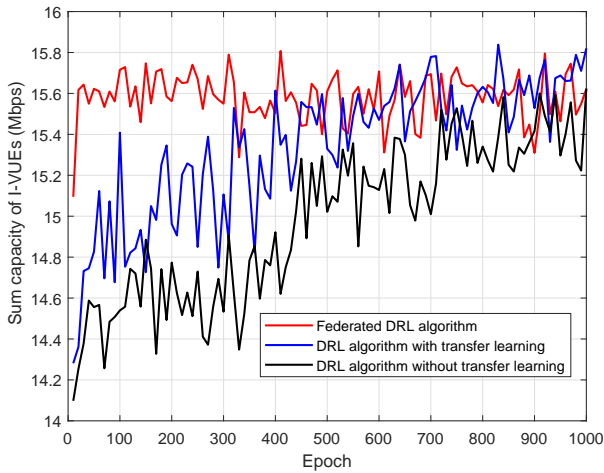}
\setlength{\belowcaptionskip}{-100pt} \vspace*{-10pt}
\caption{Sum capacity of I-VUEs in the DRL and federated DRL algorithms with a newly activated V2V pair.}
\label{NEW_RATE}
\end{figure}

\begin{figure}[!t]
\centering
\includegraphics[scale=0.6]{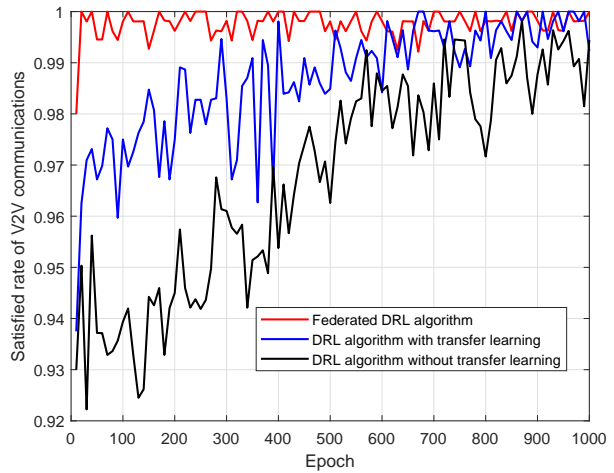}
\setlength{\belowcaptionskip}{-100pt} \vspace*{-10pt}
\caption{Satisfied rate of V2V communications in the DRL and federated DRL algorithms with a newly activated V2V pair.}
\label{NEW_PRR}
\end{figure}

Each original V2V pair selects optimal action based on its pre-trained local DRL model.
In federated DRL algorithm, the structure and weights of global model are directly downloaded to the newly activated V2V pair.
Two DRL algorithms with and without transfer learning are considered for comparison.
More specifically, the DRL algorithm with transfer learning uses local model of the closest V2V pair for this newly activated V2V pair, and then continues to train this model based on local training data.
The DRL algorithm without transfer learning trains a new model based on local scratch.
Fig. \ref{NEW_RATE} and \ref{NEW_PRR} show the learning process of above three algorithms with a newly activated V2V pair.
It is observed that in the end, the federated DRL algorithm achieves performance similar to that achieved by the other two DRL algorithms but with negligible training time.
This is because that the federated DRL trains its global model by averaging local models in the same cluster, which brings more training data of different V2V pairs
into consideration and renders corresponding global models more robust to the dynamic environment.
In addition, the federated DRL algorithm achieves stable performance while the other two DRL algorithms are more fluctuant due to exploring process and nonideal DRL model.

\section{Conclusion}
In this paper, a DRL-based transmission mode selection and resource allocation approach is designed for cellular V2X communications,
which aims to maximize the sum capacity of V2I users while guaranteeing the latency and reliability requirements of V2V pairs.
Firstly, a MDP model is built to represent considered problem, in which each V2V pair can independently select proper transmission mode, RB and power level based on local observations. Considering large continuous-value state space, a DRL-based decentralized algorithm is designed to train DRL model.
In order to train robust DRL models and improve the performance of newly activated V2V pairs,
a two-timescale federated DRL-based semi-decentralized algorithm is further developed.
Specifically, a graph-based vehicle clustering is executed on a large timescale and federated learning is conducted on a small timescale.
Simulation results have demonstrated the superiority of the proposed DRL-based algorithm with different numbers of V2V pairs and
outage thresholds, as well as the effectiveness of the proposed federated DRL algorithm for newly activated V2V pairs.

\end{document}